\documentclass[twocolumn,secnumarabic,amssymb,nobibnotes,aps,prd,showpacs,showkeys]{revtex4-1}

\usepackage{graphicx}
\newcommand{\beq}{\begin{equation}}
\newcommand{\eeq}{\end{equation}}
\newcommand{\eml}{\end{mathletters}}

\newcommand{\be}{\begin{equation}}
\newcommand{\ee}{\end{equation}}
\newcommand{\bea}{\begin{eqnarray}}
\newcommand{\eea}{\end{eqnarray}}
\newcommand{\nn}{\nonumber\\}
\newcommand{\oh}{\frac{1}{2}}

\newcommand{\la}{\langle}
\newcommand{\ra}{\rangle}

\begin{document}

\title{Pairing versus quarteting coherence length} 

\author{D.S. Delion $^{1,2,3}$ and V.V. Baran $^{1,4}$}
\affiliation{
$^1$ "Horia Hulubei" National Institute of Physics and Nuclear Engineering, \\
30 Reactorului, POB MG-6, RO-077125, Bucharest-M\u agurele, Rom\^ania \\
$^2$ Academy of Romanian Scientists, 54 Splaiul Independen\c tei RO-050085,
Bucharest, Rom\^ania \\
$^3$ Bioterra University, 81 G\^arlei RO-013724, Bucharest, Rom\^ania \\
$^4$ Department of Physics, University of Bucharest,
405 Atomi\c stilor, POB MG-11, RO-077125, Bucharest-M\u agurele, Rom\^ania}
\date{\today}

\begin{abstract}
We systematically analyze the coherence length in even-even nuclei.
The pairing coherence length in the spin-singlet channel for the effective density
dependent delta (DDD) and Gaussian interaction is estimated.
We consider in our calculations bound states as well as narrow resonances.
It turns out that the pairing gaps given by the DDD interaction are similar
to those of the Gaussian potential if one renormalizes the radial width to
the nuclear radius.
The correlations induced by the pairing interaction have in all considered cases
a long range character inside the nucleus and decrease towards the surface.
The mean coherence length is larger than the  geometrical radius for light nuclei
and approaches this value for heavy nuclei. 
The effect of the temperature and states in continuum is investigated.
Strong shell effects are evidenced, especially for protons.
We generalize this concept to quartets by considering similar relations,
but between proton and neutron pairs.
The quartet coherence length has a similar shape, but  with larger values on 
the nuclear surface.
We evidence the important role of proton-neutron correlations by estimating
the so-called alpha coherence length, which takes into account the overlap 
with  the proton-neutron part of the $\alpha$-particle wave function.
It turns out that it does not depend on the nuclear size and has a value
comparable to the free $\alpha$-particle radius.
We have shown that pairing correlations are mainly concentrated inside the nucleus,
while quarteting correlations are connected to the nuclear surface.
\end{abstract}



\maketitle

%\newpage
\setcounter{equation}{0}
\renewcommand{\theequation}{\arabic{equation}}

%\newpage 
\section{Introduction} 
\label{sec:intro} 
\setcounter{equation}{0}
\renewcommand{\theequation}{1.\arabic{equation}} 

The concept of coherence has a general character being connected to the
linear superposition of quantum states.
Two-body coherence properties in nuclear structure are directly connected to 
the properties of low-lying collective states. Collective excitations are microscopically 
described by a superposition of creation pair operators acting on the ground state,
described by a coherent state within the Random Phase Approximation (RPA).
The coherent state in this context is defined as an exponential excitation of 
products between pair operators acting on the vacuum state \cite{Rin80}.
It is well known that ground state properties of even-even nuclei 
are well reproduced by the pairing interaction \cite{Dea03,Zel04,Yos08}. 
The wave function within the Bardeen-Cooper-Schrieffer (BCS) pairing approach
is also of a coherent type, i.e. an exponential excitation of the pair creation 
operators acting on the vacuum state.

The spatial distribution of the two-particle density is very
important in understanding nuclear correlations \cite{Fer84,Hag05}.
In particular, in Ref.  \cite{Vol02} it was analyzed the relation between coherence
and chaotic properties of the nuclear pairing.
The coherence property is characterized by the so-called coherence length, defined
as the root mean square distance averaged over the density.
For superfluid nuclei this average is usually performed over the pairing density.
In Refs. \cite{Pil07,Pil10,Vin10} it was shown that this quantity is relatively large,
comparable to the nuclear size inside the nucleus and decreases beyond the nuclear surface.
This picture of an extended di-nuclear cluster can be understood in terms of the
Pauli blocking, hindering nucleons to cluster together inside the nucleus and, therefore,
the cluster loses binding and becomes larger.
It is in contrast to the $\alpha$-clustering phenomenon, which takes place
in a narrow region close to the surface area \cite{Jan83,Del10}, being connected
to the very large binding energy of an $\alpha$-particle moving in a low density region
\cite{Rop98}.
Thus, we expect that the corresponding correlation length estimated between proton 
and neutron pairs will have significantly smaller value.

The finiteness of nuclear systems also has important consequences as far as thermal properties are concerned. Pairing correlations in finite nuclei do not vanish at some critical temperature, but they slowly decrease over
several MeV \cite{Sch01, Zel03}. This can be theoretically obtained by projecting the particle number in the BCS theory \cite{Dos95}. However, hints about such behaviour can be extracted in the unprojected BCS approach from the spatial properties of the correlations.

In this paper we will perform a systematic analysis of the pairing coherence length
and a comparison to the similar quantity defined for quartets.
In Section II we give the necessary theoretical background concerning pairing
equations containing resonant states and coherence length. In Section III we perform
a systematic analysis of the coherence length and in the last Section we draw Conclusions.

\section{Theoretical background} 
\label{sec:theor} 
\setcounter{equation}{0} 
\renewcommand{\theequation}{2.\arabic{equation}} 

\subsection{Pairing equations}

In order to investigate two-body correlations we expand the wave function of $N+2$
particles in terms of the wave function of $N$ particles as follows
\bea
\label{PD}
|\Psi_{N+2}\ra
=\hat{\kappa}|\Psi_N\ra
=\sum_{\epsilon}X_{\epsilon}
\left[\hat{a}^{\dag}_{\epsilon}\otimes\hat{a}^{\dag}_{\epsilon}\right]_0|\Psi_N\ra~.
\eea
We will consider in our calculations the spherical approximation.
Thus, the operator $\hat{a}^{\dag}_{\epsilon}$ creates a single particle (sp) 
eigenstate of the spherical mean field potential with standard quantum numbers
$\epsilon\equiv(\epsilon lj)$. In the configuration representation one has
\bea
\label{psi1}
\la {\bf r,s}|\hat{a}^{\dag}_{\epsilon m}|0\ra&\equiv&\psi_{\epsilon m}({\bf r,s})=
\left[\varphi_{\epsilon}({\bf r})\otimes\chi_{\oh}({\bf s})\right]_{jm}
\nn
\varphi_{\epsilon\mu}({\bf r})&=&
\varphi_{\epsilon}(r)i^lY_{l\mu}(\hat{r})
\equiv\frac{f_{\epsilon}(r)}{r}i^lY_{l\mu}(\hat{r})~,
\eea
where  $\varphi_{\epsilon}(r)$ is the radial wave function and the rest of the
notation is standard.

The operator $\hat{\kappa}$ in Eq. (\ref{PD}) is called within the decay theory 
two-particle formation amplitude. In the absence of two-body correlations, 
when the wave functions are Slater determinants, this relation is nothing else than
the Laplace expansion of the (N+2)$\times$(N+2) normalized determinant in terms of 
N$\times$N times 2$\times$2 normalized determinants.

The most important two-body correlation beyond the mean field in even-even nuclei
is given by the pairing interaction. We will describe such systems within
the standard BCS approach, where the averaged particle number is conserved,
separately for protons and neutrons.
Thus, both wave functions in Eq. (\ref{PD}) have a BCS  ansatz and the operator $\hat{\kappa}$, 
connecting $N+2$ with $N$ systems, is called pairing density operator.
In this case the expansion coefficient
\bea
\label{X}
X_{\epsilon}&=&
\oh{\la BCS_{N+2} |\left[\hat{a}^{\dag}_{\epsilon}\otimes\hat{a}^{\dag}_{\epsilon}\right]_0|BCS_N\ra}
\nn&=&
\frac{\sqrt{2j+1}}{2}x_{\epsilon}~,
\eea
is given in terms of BCS amplitudes as follows
\bea
\label{chi}
x_{\epsilon}&\equiv&u^{(N+2)}_{\epsilon}v^{(N)}_{\epsilon}\prod_{k\ne\epsilon}
\left[u^{(N+2)}_ku^{(N)}_k+v^{(N+2)}_kv^{(N)}_k\right]
\nn
&\approx& u^{(N)}_{\epsilon}v^{(N)}_{\epsilon}
 \approx  u^{(N+2)}_{\epsilon}v^{(N+2)}_{\epsilon}~.
\eea
We will consider in our basis bound sp states with negative energy,
as well as relatively narrow sp resonances with positive energy.
Relatively narrow resonances are similar to bound states and can be normalized to unity
in the internal region, but at large distances they behave like outgoing waves
\bea
\label{sp}
\varphi_{\epsilon}(r)\rightarrow_{r\rightarrow\infty} 
M_{\epsilon}\frac{H_l^{(+)}(r)}{r}\equiv
M_{\epsilon}\frac{G_l(r)+iF_l(r)}{r}~,
\eea
in terms of spherical Hankel functions for neutrons and Coulomb-Hankel functions
for protons.
The coefficients $M_{\epsilon}$ are called scattering amplitudes and their
squared values are proportional to sp partial decay widths.

The states in continuum play an important role on pairing correlations,
especially for nuclei close to the drip lines \cite{Hag05,Ber91,Bor06,Dob96,Gra01,Ham06}. 
For nuclear structure calculations the background contribution is not relevant and only relatively
narrow resonant states are  important \cite{Bet08,Bet12}. 
A very good approximation for BCS calculations is to neglect the finite resonance width, 
i.e. to treat the resonances as bound-like states \cite{Del97}.
We label bound states by $a$ and resonances with positive energy by $r$. We treat proton and neutron pairing separately; for a given isospin index the generalized system of BCS equations for 
gap parameters $\Delta_{a}$, $\Delta_{r}$ and number of particles $N$ is
\begin{widetext}
\bea
\label{BCS}
\Delta_{a}&=&\sum_{a'}\left(j_{a'}+\frac{1}{2}\right)V_{a,a'}
\frac{\Delta_{a'}}{2\sqrt{(\epsilon_{a'}-\lambda)^{2}+\Delta_{a'}^{2}}}+
\sum_{r}\left(j_{r}+\frac{1}{2}\right)V_{a,r}
\frac{\Delta_{r}}{2\sqrt{(\epsilon_{r}-\lambda)^{2}+\Delta_{r
}^{2}}}~,
\nn
\Delta_{r}&=&\sum_{a'}\left(j_{a'}+\frac{1}{2}\right)V_{r,a'}
\frac{\Delta_{a'}}{2\sqrt{(\epsilon_{a'}-\lambda)^{2}+\Delta_{a'}^{2}}}+
\sum_{r'}\left(j_{r'}+\frac{1}{2}\right)V_{r,r'}
\frac{\Delta_{r'}}{2\sqrt{(\epsilon_{r'}-\lambda)^{2}+\Delta_{r'
}^{2}}}~,
\nn
N&=&\sum_{a}\left(j_{a}+\oh\right)\left(1-\frac{\epsilon_{a}-\lambda}{\sqrt{(\epsilon_{a}-\lambda)^{2}+
\Delta_{a}^{2}}}\right)+\sum_{r}(j_{r}+\oh)\left(1-\frac{\epsilon_{r}-\lambda}{\sqrt{(\epsilon_{r}-\lambda)^{2}+
\Delta_{r}^{2}}}\right)~,
\eea
\end{widetext}
where $\lambda$ is the chemical potential and the potential matrix elements $V_{\alpha,\beta}$
are computed according to Eq. (2.6) of \cite{Del95}.

We will investigate pairing in excited nuclei by using the temperature-dependent equations
with anomalous and normal densities, respectively
\bea
\la a_{\epsilon}a_{\bar{\epsilon}}\ra&=&u_{\epsilon}v_{\epsilon}\text{tanh}\frac{\beta E_{\epsilon}}{2}
\nn
\la a_{\epsilon}^\dag a_{\epsilon} \ra&=&v_{\epsilon}^2 +(u_{\epsilon}^2-v_{\epsilon}^2) 
/(\text{e}^{\beta E_{\epsilon}}+1)~.
\eea

\subsection{Pairing coherence length}

The two-body operator entering  the pairing density (\ref{PD})
can be written in the configuration representation.
By using the recoupling from j-j to the L-S scheme, one obtains spin-singlet
and spin-triplet components. Our calculations have shown that the largest contribution
is given by the spin-singlet component, given  the following expression
\bea
\label{singlet}
\kappa({\bf r_1,r_2})&=&\sum_{\epsilon}z_{\epsilon}
\left[\varphi_{\epsilon}({\bf r}_1)\otimes\varphi_{\epsilon}({\bf r}_2)\right]_0
\nn&=&
\sum_{\epsilon}z_{\epsilon}\frac{f_{\epsilon}(r_1)f_{\epsilon}(r_2)}{r_1r_2}
{\cal Y}_l(\cos\theta)~,
\eea
in terms of two-particle azimuthal harmonics
\bea
\label{azimut}
{\cal Y}_{l}(\cos\theta)&=&
\left[i^{l}Y_{l}(\hat{r}_1)\otimes i^{l}Y_{l}(\hat{r}_2)\right]_0
\nn&=&
\frac{\sqrt{2l+1}}{4\pi}P_{l}(\cos\theta)~,
\eea
where $\theta$ is the angle between particle radii and
the expansion coefficient is given by
\bea
z_{\epsilon}&=&x_{\epsilon}\sqrt{j+\oh}
\la (ll)0,\left(\oh\oh\right)0;0|\left(l\oh\right)j,\left(l\oh\right)j;0\ra~,
\nn
\eea
in terms of LS-jj recoupling brackets.
By expanding  the sp wave function with respect to the harmonic oscillator (ho) basis
\bea
\varphi_{\epsilon\mu}({\bf r})&=&\sum_{n}c_{n\epsilon}\phi^{(\beta)}_{nl\mu}({\bf r})
\nn
\phi^{(\beta)}_{nl\mu}({\bf r})&=&\phi^{(\beta)}_{nl}(r)i^lY_{l\mu}(\hat{r})~,
\eea
where $\beta=M_N\omega/\hbar$ is the standard ho parameter,
and by using  the Talmi-Moshinski transformation to relative ${\bf r=r_1-r_2}$ and 
center of mass (c.o.m.) coordinate ${\bf R=(r_1+r_2)}/2$
one obtains the following expansion
\bea
\label{kappa}
&&\kappa(r,R,\theta)=\sum_{\lambda} 
f_{\lambda}(r,R){\cal Y}_{\lambda}(\cos\theta)~,
\eea
with expansion coefficients given by
\bea
\label{flam}
f_{\lambda}(r,R)&=&\sum_{nN}
{\cal G}_{nN\lambda}\phi^{(\beta/2)}_{n \lambda}(r)\ \Phi^{(2\beta)}_{N\lambda}(R)~,
\eea
where
\bea
\label{G}
{\cal G}_{nN\lambda}&\equiv&\sum_{\epsilon}z_{\epsilon}\sum_{n_1n_2}c_{n_1\epsilon}c_{n_2\epsilon}
\la n\lambda N\lambda;0|n_1 l n_2 l;0\ra~.
\nn
\eea
Here, the bracket denotes the standard Talmi-Moshinsky recoupling coefficient.
By averaging over the angle $\theta$ we get
\bea
\label{average}
\bar{\kappa}^2(r,R)&=&\frac{1}{2}\int_{-1}^1 \kappa^2(r,R,\theta){\rm d}\cos\theta
\nn&=&
\frac{1}{(4\pi)^2}\sum_{\lambda} f_\lambda^2(r,R)~.
\eea
The coherence length is defined as follows
\bea
\label{cohlen}
\xi(R)&=&\sqrt{\frac{I^{(2)}(R)}{I^{(1)}(R)}}
\nn
&\equiv&\sqrt{\int_0^{\infty}{\rm d}r\ r^2\ w(r,R)}~,
\eea
in terms of the integrals
\bea
\label{int}
&&I^{(p)}(R)\equiv\int_0^{\infty} \text{d}r\ r^{2p}\ \bar{\kappa}^2(r,R)
\nn&=&
\sum_{\lambda NN'}\Phi^{(2\beta)}_{N\lambda}(R)\Phi^{(2\beta)}_{N'\lambda}(R)
\nn&\times&
\sum_{nn'}{\cal G}_{nN\lambda}{\cal G}_{n'N'\lambda}
\int_0^{\infty} \text{d}r\ r^{2p}\phi^{(\beta/2)}_{n\lambda}(r)\phi^{(\beta/2)}_{n'\lambda}(r)~.
\nn
\eea
Let us finally mention that the quantity $x_{\epsilon}$, defined by Eq. (\ref{chi}),
is also called "anomalous" density, while the quantity
\bea
y_{\epsilon}=v^2_{\epsilon}~.
\eea
is called "normal" density. Therefore $\kappa$, defined by Eq. (\ref{singlet}),
can be called "anomalous" coherence length, while a similar quantity $\kappa_0$ defined
by using the "normal" density is called "normal" coherence length.

\subsection{Quarteting correlations}

We will investigate quarteting correlations in medium and heavy $\alpha$-decaying nuclei,
where the valence protons and neutrons occupy different major shells.
The standard assumption to build a quartet from two protons and two neutrons in such nuclei
is to consider proton and neutron pairing separately \cite{Man60,San62}.
Therefore the system of $N_{\pi}+2,N_{\nu}+2$ nucleons can be expressed in terms of 
$N_{\pi},N_{\nu}$ nucleons in a factorized way as follows
\bea
\label{PD4}
|\Psi_{N_{\pi}+2,N_{\nu}+2}\ra
=\hat{\kappa}_{\pi}\hat{\kappa}_{\nu}|\Psi_{N_{\pi}N_{\nu}}\ra~,
\eea
where $\kappa_{\tau}$ is defined by Eq. (\ref{PD}).
Thus, the quartet wave function is a product between proton and neutron two-body
wave functions  (\ref{kappa}). Anyway, calculations in infinite nuclear matter
suggest that $\alpha$-clusters can occur only at relative small nuclear densities compared
to the equilibrium value and the proton-neutron correlations play an important role \cite{Rop98}.
Thus, an $\alpha$-particle can be formed only in the surface region where the nuclear
density diminishes and proton-neutron correlations become relevant. This situation
can be simulated by a proper modification of the single particle mean field
by adding a gaussian interaction in the surface region \cite{Del13} and still by 
keeping the factorized ansatz (\ref{PD4}). This can expain why an $\alpha$-particle
can be formed from two protons and two neutrons lying in different major shells.
This additional ansatz of the single particle mean field was recently confirmed by
microscopic calculations \cite{Rop14} and fission-like theory \cite{Mir15}.
Anyway, this modification is important in order to reproduce the absolute value of
the half-life, but has a minor influence on the coherence length. 

In order to describe quartets we introduce the relative and c.o.m. coordinates
for proton, neutron and proton-neutron systems, respectively:
\bea
&&{\bf r}_{\pi}={\bf r}_1-{\bf r}_2~,~~~{\bf R}_{\pi}=\frac{{\bf r}_1+{\bf r}_2}{2}
\nn
&&{\bf r}_{\nu}={\bf r}_3-{\bf r}_4~,~~~{\bf R}_{\nu}=\frac{{\bf r}_3+{\bf r}_4}{2}
\nn
&&{\bf r}_{\alpha}={\bf R}_{\pi}-{\bf R}_{\nu}~,~~~
{\bf R}_{\alpha}=\frac{{\bf R}_{\pi}+{\bf R}_{\nu}}{2}~,
\eea
where we labeled by $1,~2$ proton and by $3,~4$ neutron coodinates.
The internal $\alpha$-particle wave function is given by the product
between the lowest proton, neutron and proton-neutron ho orbitals
\bea
\label{psialpha}
\psi_{\alpha}=\phi^{(\beta_{\alpha}/2)}_{00}(r_{\pi})\phi^{(\beta_{\alpha}/2)}_{00}(r_{\nu})
\phi^{(\beta_{\alpha})}_{00}(r_{\alpha})~,
\eea
where $\beta_{\alpha}\approx$ 0.5 $fm^{-2}$ is the free $\alpha$-particle ho 
parameter measured by electron scattering experiments \cite{Man60}.
This parameter is about 2-3 times larger than the similar sp ho parameter 
in heavy $\alpha$-emitters, due to the fact that $\alpha$-particle is a very bound object.

We will describe quarteting correlations between proton and neutron pairs
by overlapping the relative coordinates to the corresponding components of the
$\alpha$-particle wave function (\ref{psialpha}).
We will proceed in two steps.

\vskip5mm
A. {\it Quarteting correlation length}
\vskip5mm

Let us first consider only the overlap with respect to proton and neutron relative
coordinates $r_{\pi},~r_{\nu}$, by keeping free the internal proton-neutron coordinate $r_{\alpha}$.
Thus, we consider independent from each other proton and neutron pairs by
neglecting proton-neutron correlations.
Therefore we can define the quarteting density in analogy to the pairing density, but between
the proton and neutron pairs (instead of fermions):
\bea
\label{kappaq}
\kappa_q({\bf R}_\pi,{\bf R}_\nu)&=&
\la\kappa_{\pi}({\bf r_1,r_2})|\phi^{(\beta_{\alpha}/2)}_{00}(r_{\pi})\ra
\nn&\times&
\la\kappa_{\nu}({\bf r_3,r_4})|\phi^{(\beta_{\alpha}/2)}_{00}(r_{\nu})\ra~.
\eea
By recoupling the product between proton and neutron pairs (\ref{kappa}) to the relative
and c.o.m. pair coordinates one obtains for the leading monopole component the following relation
%\begin{widetext}
\bea
&&\kappa_q(r_\alpha,R_\alpha)\approx\kappa^{(0)}_q(r_\alpha,R_\alpha)=
\sum_{N_\pi,N_\nu}G_\pi(N_\pi)\ G_\nu(N_\nu)
\nn&\times&
\sum_{n_{\alpha}}\la n_{\alpha}0N_{\alpha}0;0|N_\pi 0N_\nu 0;0\ra 
\phi^{(\beta)}_{n_{\alpha}0}(r_\alpha) \phi^{(4\beta)}_{N_{\alpha}0}(R_\alpha)~,
\nn
\eea
%\end{widetext}
in terms of Moshinsky brackets and the proton/neutron monopole G-coefficients (\ref{G})
\bea
G_{\tau}(N_{\tau})&=&\sum_{n_{\tau}}{\cal G}_{n_{\tau}N_{\tau}0}
\la \phi^{(\beta/2)}_{n_{\tau}0}(r_{\tau})|\phi^{(\beta_{\alpha}/2)}_{00}(r_{\tau})\ra
\nn
\tau&=&\pi,\nu~.
\eea
It does not depend on angles and therefore one can define the quarteting coherence length 
$\xi_q(R_{\alpha})$ without any additional angular average (\ref{average}) by using in
Eqs. (\ref{cohlen}) and (\ref{int}) the quarteting density squared $\kappa^2_q(r_{\alpha},R_{\alpha})$.

\vskip5mm
B. {\it Alpha coherence length}
\vskip5mm

The next step is to consider proton-neutron correlations.
They are described by the corresponding part in the $\alpha$-particle wave function 
(\ref{psialpha}) given by $\phi^{(\beta_{\alpha})}_{00}(r_{\alpha})$.
In order to account for the narrow proton-neutron spatial distribution 
in the free $\alpha$-particle one defines the so-called
alpha coherence length $\xi_{\alpha}(R_{\alpha})$ by using the alpha density
\bea
\label{kappaa}
\kappa_{\alpha}(r_{\alpha},R_{\alpha})=
\kappa_q(r_{\alpha},R_{\alpha})\phi_{00}^{(\beta_{\alpha})}(r_{\alpha})~,
\eea
in performing the integrals (\ref{int}).

Let us finally mention that the integral of the alpha density
over the relative proton-neutron coordinate
\bea
\label{ampl}
{\cal F}(R_{\alpha})=\int_0^{\infty}\kappa_{\alpha}(r_{\alpha},R_{\alpha})r^2_{\alpha}dr_{\alpha}~,
\eea
defines the formation amplitude and its square describes the probability to find an $\alpha$-particle
in the quartet wave function \cite{Man60,Del10}.

\begin{figure}[ht] 
\begin{center} 
\includegraphics[width=9cm]{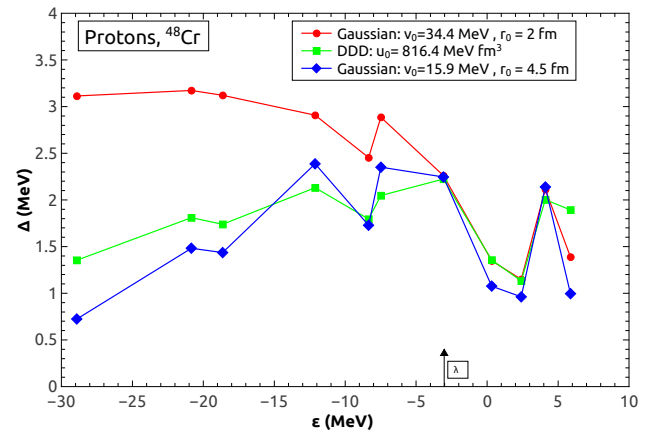} 
\vspace{-8mm}
\caption{
Paring gap defined by the first two lines of Eq. (\ref{BCS}) versus $\epsilon$ 
in $^{48}$Cr for DDD potential (squares) and Gaussian potentials
with $r_0$= 2 fm (circles) and $r_0=R_N$ (diamonds).
}
\label{fig01}
\end{center} 
\end{figure}

\section{Numerical application} 
\label{sec:appl} 
\setcounter{equation}{0}
\renewcommand{\theequation}{3.\arabic{equation}} 

We analyzed all even-even nuclei with $20<Z<100$ and known experimental pairing gaps, 
determined by the binding energies of neighbouring nuclei \cite{Mol95}.

\begin{widetext}
\begin{center}
\begin{table}
\caption{\rm Proton quantum numbers, sp spectrum, decay widths and gap parameters 
for the Gaussian, renormalized Gaussian and DDD interactions in $^{48}$Cr, given by 
the diagonalization of the  Woods-Saxon mean field with universal parametrisation \cite{Dud80}.
\\}
\label{tab1}
\begin{tabular}{|c|c|c|c|c|c|c|c|}
\hline
No. & $l$ & $2j$ & $\epsilon$ (MeV) & $\Gamma$ (MeV) & $\Delta_{ 2\text{fm}}$(MeV) & $\Delta_{ 4.5\text{fm}}$(MeV)& $\Delta_{ \text{DDD}}$(MeV) \cr
\hline
 1 & 0 & 1 &-28.911 &    -   & 3.114 & 1.354 &  0.724 \\
 2 & 1 & 3 &-20.837 &    -   & 3.173 & 1.810 &  1.482 \\ 
 3 & 1 & 1 &-18.638 &    -   & 3.121 & 1.739 &  1.436 \\
 4 & 2 & 5 &-12.118 &    -   & 2.908 & 2.131 &  2.387 \\
 5 & 0 & 1 & -8.349 &    -   & 2.454 & 1.795 &  1.728 \\
 6 & 2 & 3 & -7.488 &    -   & 2.886 & 2.047 &  2.351 \\
 7 & 3 & 7 & -3.079 &    -   &  .261 & 2.224 &  2.246 \\
 8 & 1 & 3 &  0.322 &  0.000 & 1.349 & 1.356 &  1.076 \\
 9 & 1 & 1 &  2.403 &  0.046 & 1.149 & 1.133 &  0.962 \\
10 & 3 & 5 &  4.101 &  0.024 & 2.114 & 2.003 &  2.139 \\
11 & 4 & 9 &  5.874 &  0.055 & 1.389 & 1.893 &  0.996 \\
\hline

\end{tabular}
\end{table}
\end{center}
\end{widetext}

For the nuclear mean field we used a standard Woods-Saxon potential
with universal  parametrization \cite{Dud80}.
We considered in our sp basis all bound states and resonances in continuum 
up to $e_{max}=$ 10 MeV with a sp decay width  $\Gamma\leq$ 1 MeV.
As an example we give in Table \ref{tab1} the proton sp spectrum for $^{48}$Cr.
Here, there are given level number, angular momentum, twice the total spin,
sp energy, decay width of sp states in continuum and
pairing gaps for the interactions considered below.

\begin{figure}[ht] 
\begin{center} 
\includegraphics[width=9cm]{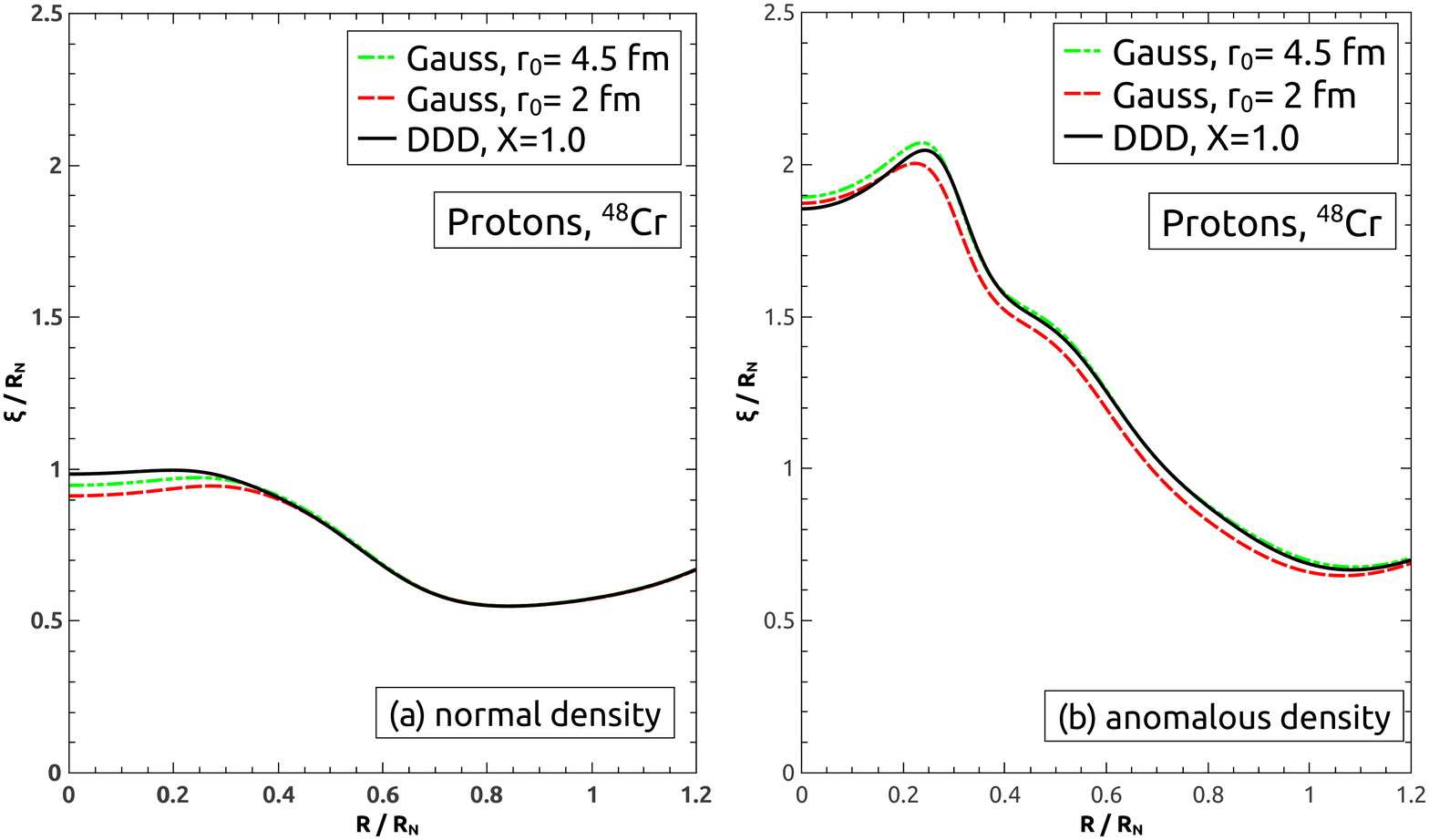} 
\vspace{-8mm}
\caption{
Proton coherence length divided by geometrical radius versus c.o.m. radius 
in $^{48}$Cr computed with "normal" (a) and "anomalous" densities (b)
for DDD potential (solid line) and Gaussian potentials
with $r_0$= 2 fm (long dashes) and $r_0=R_N$ (short dashes).
}
\label{fig02}
\end{center}
\end{figure}

We solved the BCS equations (\ref{BCS}) separately for protons and neutrons
with two widely used types of nucleon-nucleon pairing interactions:

\vskip5mm
I. {\it Gaussian interaction}
\vskip5mm

It is defined by the following ansatz:
\bea
\label{vr12}
v(r_{12})&=&-v_0e^{-\left[r_{12}/r_0\right]^2}~,
\eea
depending on the relative radius $r_{12}$.
Here, the width parameter $r_0$=2 fm corresponds to the spin-singlet "bare" value in the free space.
The corresponding value of the effective potential strength $v_0$ is determined 
by the gap parameter at the Fermi level, which should be equal to the experimental value.

\vskip5mm
II. {\it Density dependent delta (DDD) interaction}
\vskip5mm

It is known that the strength of the effective pairing interaction depends upon  
the local density \cite{Ber91,Bor06}, given by the following phenomenological ansatz \cite{Dob96}
\bea
\label{dens}
v({\bf r,r'})=u_0\delta({\bf r-r'})
\left\lbrace1-X\left[\frac{\rho_N({\bf r})}{\rho_N^{(0)}}\right]^{\gamma}\right\rbrace~,
\eea
in terms of the nuclear density $\rho_N$.
The value $X$=1 corresponds to the surface DDD interaction.

As an example, in Fig. \ref{fig01} we plotted the pairing gap (\ref{BCS}) versus
sp energy for $^{48}$Cr, given in Table \ref{tab1}.
Here, circles correspond to the Gaussian interaction in the free space with $r_0$=2 fm.
Notice large values for states below the Fermi level. The gaps given by DDD interaction
with $X=\gamma$=1 are plotted by squares and the values below the Fermi level are significantly 
smaller that the Fermi gap. 

\begin{figure}[ht] 
\begin{center} 
\includegraphics[width=9cm]{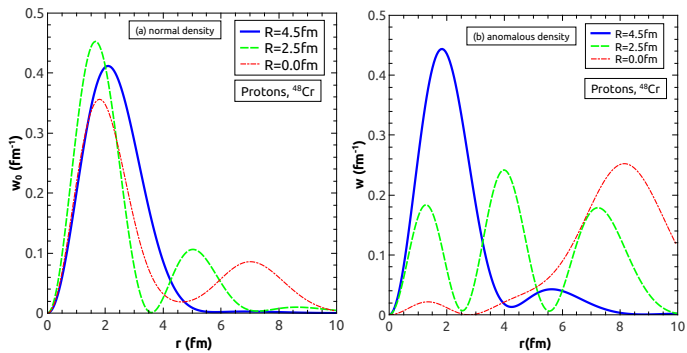} 
%\vspace{-8mm}
\caption{
The integrand of the proton coherence length versus the relative radius
radius in $^{48}$Cr, computed with "normal" (a) and "anomalous" densities (b)
for different c.o.m. radii. Here, we used the Gaussian interaction withn $r_0$=2 fm.
}
\label{fig03}
\end{center} 
\end{figure}

It is interesting to point out that a very similar behaviour has the Gaussian interaction
where the width parameter is renormalized to the geometrical nuclear radius (in fm)
$r_0=R_N=1.2A^{1/3}$.
A di-nuclear cluster inside nuclear matter has different properties with respect to
the free space. It considerably loses the binding property due to the Pauli blocking,
becoming larger and therefore the effective pairing interaction has a more extended shape.
Above the Fermi sea we obtained similar values in all cases.

In Fig. \ref{fig02} (a) we plotted the proton coherence length given by 
Eq. (\ref{cohlen}) divided by the nuclear radius $R_N$, as a function of the
ratio between c.o.m. and nuclear radius $R/R_N$ in $^{48}$Cr. Here we used the "normal" density
while in  Fig. \ref{fig02} (b)  we used the "anomalous" density.
Notice that that all cases, plotted by different symbols explained in caption,
have very similar shapes. Thus, the coherence length is not sensitive to the
radial shape of the interaction.
The "normal" coherence length is equal to the nuclear radius in the internal region and diminishes by a factor 0.5 on the surface.
The "anomalous" coherence length has a similar shape, but with twice larger internal value.
This picture is very different from the dependence of the two-body wave function versus
c.o.m. radius, which is peaked on the nuclear surface \cite{Del95}.

\begin{figure}[ht] 
\begin{center} 
\includegraphics[width=9cm]{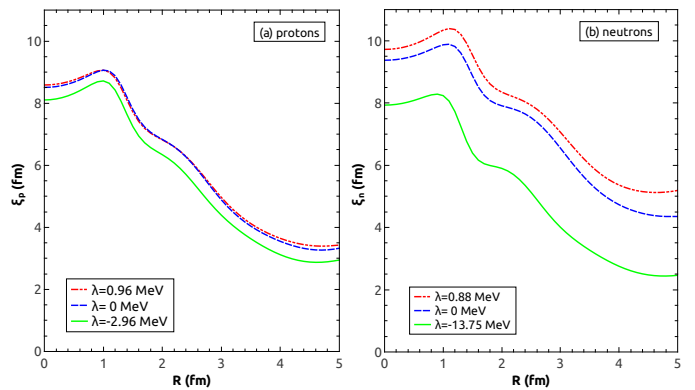} 
%\vspace{-8mm}
\caption{
(a) Proton coherence length versus c.o.m. radius for different chemical potentials
$\lambda$=-2.96 MeV (solid line), 0 MeV (long dashes) and 0.96 MeV (short dashes)
in $^{48}$Cr.\\
(b) Neutron coherence length versus c.o.m. radius for different chemical potentials
$\lambda$=-13.75 MeV (solid line), 0 MeV (long dashes) and 0.88 MeV (short dashes)
in $^{48}$Cr.
}
\label{fig04}
\end{center} 
\end{figure}

\begin{figure}[ht] 
\begin{center} 
\includegraphics[width=9cm]{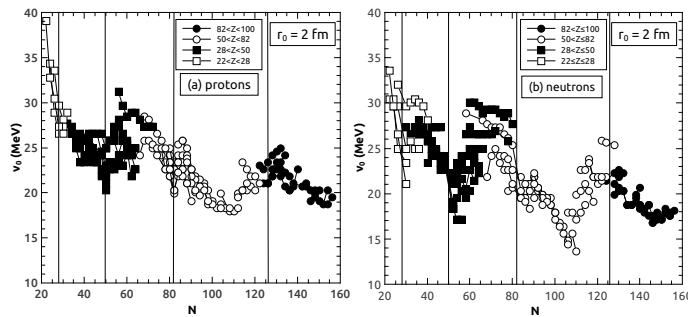} 
%\vspace{-8mm}
\caption{
Strength parameter of the Gaussian interaction, corresponding to $r_0$=2 fm, 
versus neutron number for proton (a) and neutron systems (b).
}
\label{fig05}
\end{center} 
\end{figure}

In order to better understand the behaviour of the coherence length we plotted
in Fig. \ref{fig03} (a) the integrand of the "normal" correlation length $w_0(r,R)$,
given by the second line of Eq. (\ref{cohlen}), versus the relative radius $r$
for three values of the c.o.m. radius $R$=4.5 fm (solid line), 2 fm (long dashes) and
0 fm (short dashes). Here, we used the "bare" version of the Gaussian interaction.
Notice that the three curves have a similar shape, strongly peaked 
around 2 fm. We obtain completely different plots for the integrand of the "anomalous"
coherence length $w(r,R)$. They are given in Fig. \ref{fig03} (b). 
The distribution corresponding to the c.o.m. radius on surface $R$=4.5 fm (solid line)
is peaked around the free singlet value of the Gaussian width i.e. $r$=2 fm.
On the contrary, the distribution corresponding to a smaller radius $R$=2.5 (long dashes)
is peaked around a much larger value $r$=7 fm.

Our conclusions are in agreement with Ref. \cite{Pil07}, where in Fig. 5 the 
"anomalous" coherence length of the pairing interaction was estimated within 
the more sophisticated Hartree-Fock-Bogoljubov (HFB) approach, by using the Gogny 
force for Ni isotopes.
The shape is similar, predicting a mean coherence length of about 6 fm in the internal region
and decreasing as one approaches the nuclear surface and reaching the value of 2 fm
just outside the nucleus.

\begin{figure}[ht] 
\begin{center} 
\includegraphics[width=9cm]{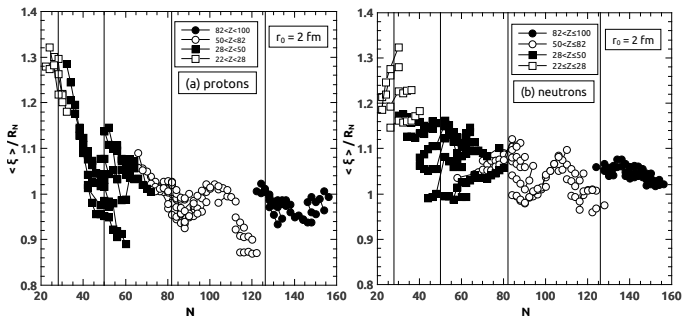} 
%\vspace{-8mm}
\caption{
Ratio $\la\xi\ra/R_N$, corresponding to a Gaussian interaction with $r_0$=2 fm, 
versus neutron number for proton (a) and neutron systems (b).
}
\label{fig06}
\end{center} 
\end{figure}
 
\begin{figure}[ht] 
\begin{center} 
\includegraphics[width=9cm]{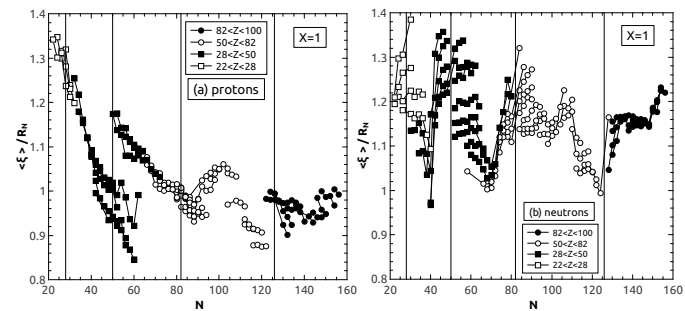} 
%\vspace{-8mm}
\caption{
Ratio $\la\xi\ra/R_N$, corresponding to the DDD interaction (\ref{dens}) with $X=\gamma=1$, 
versus neutron number for proton (a) and neutron systems (b).
}
\label{fig07}
\end{center} 
\end{figure}

Most of the exotic nuclei close to the drip lines have the last nucleon in continuum.
Therefore we investigated the dependence of the coherence length on the Fermi level,
by changing the real part of the Woods-Saxon potential.
We plotted in Fig. \ref{fig04} (a) the proton coherence length versus c.o.m. radius in $^{48}$Cr,
for different values of the chemical potential.
One sees that it increases by increasing the chemical potential. 
This effect is stronger for neutrons, as seen in Fig.  \ref{fig04} (b),
due to the absence of the Coulomb barrier. Therefore, in exotic nuclei close
to drip lines the nucleons become more correlated.

\begin{figure}[ht] 
\begin{center}
\includegraphics[width=9cm]{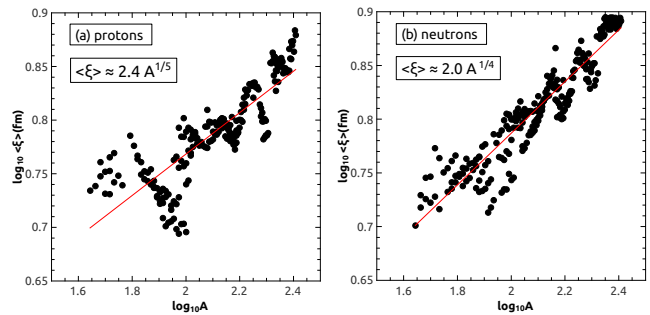} 
%\vspace{-8mm}
\caption{
Logarithm of the ratio $\la\xi\ra/R_N$ versus logarithm of the mass number
for protons (a) and neutrons (b).
}
\label{fig08}
\end{center} 
\end{figure}

Then we performed a systematic analysis of the "anomalous" coherence length 
(by simply calling it coherence length) for even-even nuclei with $20<Z<100$. 

In Fig. \ref{fig05} (a) we plotted the effective strength $v_0$ as a function of the neutron number
for protons corresponding to Gaussian interaction with $r_0$=2 fm.
The isotope chains are connected by solid lines and magic numbers are indicated by vertical lines.
Different regions are plotted by open squares ($20<Z<28$), filled squares ($28<Z<50$),
open circles ($50<Z<82$) and filled circles ($82<Z<100$).
As a general trend we remark a strong decreasing behaviour with the increase 
of the neutron number. We notice a remarkable feature, namely it has
almost the singlet "bare" value in the free space $v_0\sim$ 35 MeV for very light nuclei. 
The strength strongly decreases up to $v_0\sim$ 20 MeV for heavy nuclei,
except the regions around magic numbers.
In Fig. \ref{fig05} (b) we give a similar plot for neutrons. Notice that in this
case shell effects are stronger.

\begin{figure}[ht] 
\begin{center} 
\includegraphics[width=9cm]{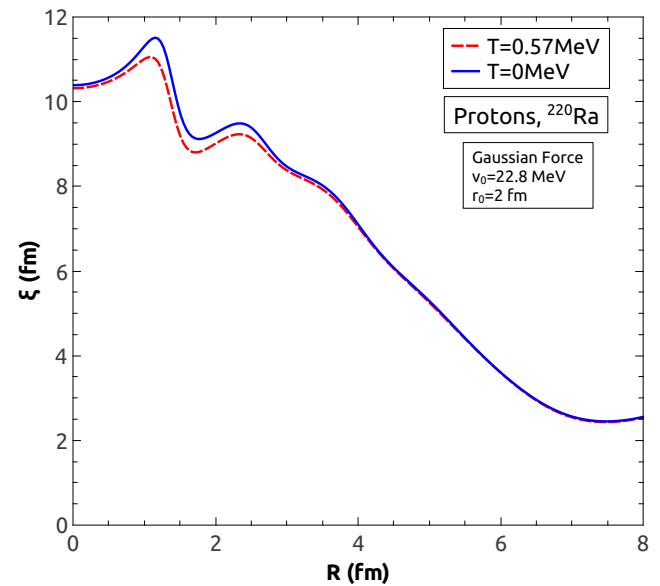} 
%\vspace{-8mm}
\caption{
Proton coherence length versus c.o.m. radius 
in $^{220}$Ra computed for T=0 (solid line), T=0.5675 MeV $\lesssim T_c$ (long dashes), for a Gaussian potential with $r_0$= 2 fm.
}
\label{fig09}
\end{center} 
\end{figure}

%\newpage
\begin{figure}[ht] 
\begin{center} 
\includegraphics[width=9cm]{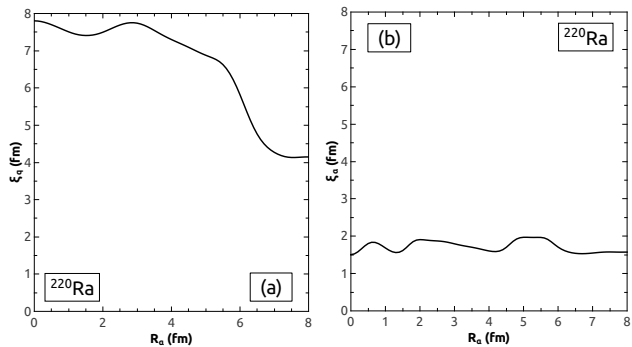}
%\vspace{-8mm}
\caption{
(a) Quarteting coherence length in $^{220}$Ra versus c.o.m. radius. \\
(b) Same as in (a) but for alpha coherence length.
}
\label{fig10}
\end{center} 
\end{figure}

%\newpage
\begin{figure}[ht] 
\begin{center} 
\includegraphics[width=9cm]{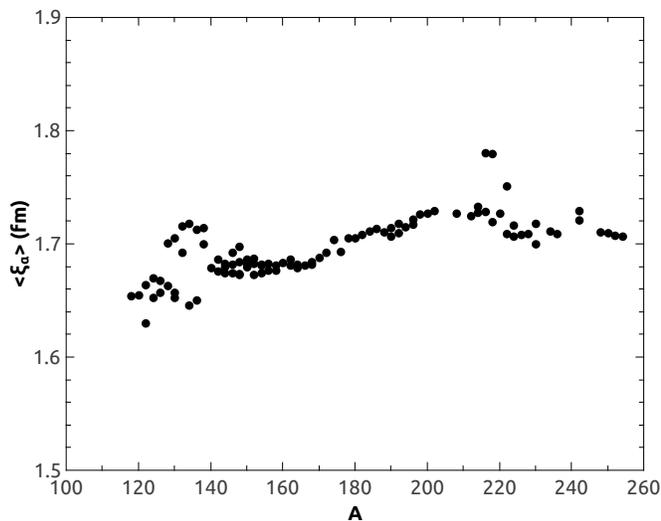} 
%\vspace{-8mm}
\caption{
Averaged alpha coherence length versus mass number.
}
\label{fig11}
\end{center} 
\end{figure}

In Fig. \ref{fig06} (a) we analyzed the mean coherence length $\la\xi\ra$ 
for protons, corresponding to the Gaussian interaction with the free value
of the width parameter $r_0$=2 fm as a function of neutrons. 
The ratio of this quantity to the nuclear radius decreases from 1.4 for light nuclei
up to around unity for heavy nuclei.
In Fig. \ref{fig06} (b) we give similar results for neutrons. As a general
trend, the coherence length is larger for neutrons due to the absence of the Coulomb
barrier, but the shell effects are stronger for protons.

We then investigated the density dependent pairing interaction given by Eq. (\ref{dens})
with $X=\gamma$=1 in Fig. \ref{fig07}.
It turns out that the ratio $\la\xi\ra/R_N$ has similar gross features, but with
more pronounced shell oscillations. The fact that the coherence length for neutrons
is larger is confirmed.
It is interesting to notice the linear correlation between $log_{10}\la\xi\ra$
and $log_{10}A$, plotted in Fig. \ref{fig08} for the Gaussian pairing interaction with $r_0$= 2 fm.

%%%%%
In order to investigate the behaviour of the coherence length for excited
states, in Fig. \ref{fig09} we analyzed the role of the temperature.
Firstly, we give for $^{220}$Ra the coherence length versus the pair
c.o.m. radius for $T=0$ and just below the 'critical' temperature $T_c
\approx 0.57$ MeV (here the gap decreases below $10^{-3}$ MeV). The
pairing coherence length shows very little change in shape up to $T_c$.
The strongest variation appears in the internal region, while on the
surface, where the pairs are strongly coupled \cite{Pil07}, there is
indeed almost no change. As a measure of the pairing correlations, the
coherence lenght would appear to indicate a gradual transition to the
normal state with increasing temperature, as its behaviour is similar to
that of the pairing gap in the particle number conserving case
\cite{Liu14, Dos95}.
%%%%%
%\newpage
\begin{figure}[ht] 
\begin{center} 
\includegraphics[width=9cm]{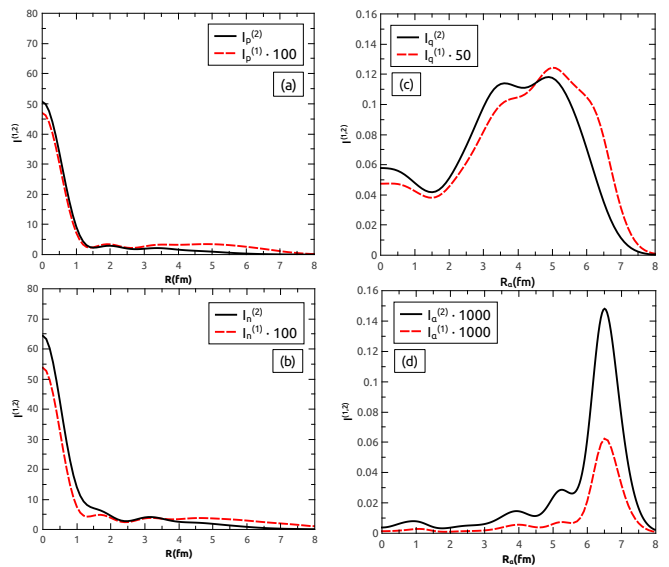} 
%\vspace{-8mm}
\caption{
The two terms $I^{(p)}(R)$, p=2 (solid line) p=1 (dashed line) given by Eq,. (\ref{int}),
defining the pairing coherence length for protons (a), neutrons (b),
quarteting coherence length (c) and alpha coherence length (d) versus the c.o.m. radius.
}
\label{fig12}
\end{center} 
\end{figure}

Our purpose is to compare the pairing and quarteting coherence lengths.
First we analyzed the quarteting coherence length, by using the quarteting
density (\ref{kappaq}), for the $\alpha$-emitter $^{220}$Ra
as a function of c.o.m. radius in Fig. \ref{fig10} (a).
One notices a similar qualitative behaviour compared to the pairing coherence length,
but the absolute values are larger on the nuclear surface.
Our calculations have shown  that the temperature practically does not change this dependence.

It turns out that the proton-neutron correlations, given by the overlap with the corresponding
proton-neutron part of the $\alpha$-particle wave function (\ref{kappaa}),
completely change this picture.
One sees from Fig. \ref{fig10} (b), where we plotted the alpha correlation length versus c.o.m. radius,
that the values oscillate around the value of the geometrical radius of the $\alpha$-particle $R_{\alpha}$.
Thus, our analysis confirm the crucial role played by proton-neutron
correlations in the formation of the $\alpha$-particle.
Finally, in Fig. \ref{fig11} we plotted the mean value of the alpha
coherence length for even-even $\alpha$-emitters above $A=100$. It has
a quasi-constant value around 1.7 fm. Small local maxima correspond to regions
above double magic nuclei $^{132}$Sn and $^{208}$Pb.

In order to better understand the difference between pairing and quarteting correlations
we plotted in Fig. \ref{fig12} the two terms $I^{(p)}$, p=2 (solid line) and p=1
(dashed line) given by Eq. (\ref{int}) versus the c.o.m. radius.
The two terms reach their maximal values for the pairing case (left panels) at
R=0, while for the quarteting case (right panels) the maxima are
centered around the surface region. 
The pairing coherence length for protons (a) and neutrons (b) is given by
the ratio between solid and dashed curves which obviously decreases with increasing c.o.m. radius.
Quarteting coherence length is given by the ratio between solid and dashed lines
in Fig. \ref{fig12} (c) which have slighly shifted broad maxima located
below the nuclear surface. Although the two terms have completely different
shapes compared to the pairing case, their ratio plotted in Fig. \ref{fig10} (a)
is also a decreasing function with respect to the c.o.m. radius.

The alpha coherence length, given by the ratio
of the two curves in Fig. \ref{fig12} (d), deserves special attention. 
These curves have very narrow maxima centered at the same point on the nuclear surface.
Moreover, it turns out that the two curves are almost proportional and therefore their
ratio leads to the quasiconstant value in Fig. \ref{fig10} (b),
close the $\alpha$-particle geometrical
radius $R^{(0)}_{\alpha}=1.2~4^{1/3}\approx$ 1.9 fm.
Notice that the shape of the curves in Fig. \ref{fig12} (d), peaked on the nuclear surface,
is similar to the standard $\alpha$-particle formation probability,
given by the integral (\ref{ampl}) squared \cite{Del10,Del13}. 

%\newpage
\section{Conclusions} 
\label{sec:concl} 
\setcounter{equation}{0} 
\renewcommand{\theequation}{4.\arabic{equation}} 

In conclusion, we have performed in this paper a systematic analysis of the
pairing coherence length in the spin-singlet channel for various types of pairing interaction.
We compared the DDD potential to the Gaussian interaction.
We considered in our calculations bound states as well as narrow resonances.

As a very important conclusion we have shown that, by considering the singlet "bare" 
value of the width parameter $r_0$=2 fm, the strength parameter reproducing the gap parameter
for light nuclei is  close to the singlet value in the free space $v_0\sim$ 35 MeV and decreases 
up to $v_0\sim$ 20 MeV for heavy nuclei.
We have shown that the "renormalized" Gaussian interaction with a larger width parameter 
than its free value $r_0$=2 fm (equal to the nuclear radius) has similar properties 
to the commonly used density dependent pairing potential.

It turns out that the pairing coherence length has similar properties for all considered
interactions. It is larger than the geometrical radius for light nuclei and
approaches this value for heavy nuclei. Our analysis evidenced strong shell effects.

%%%%%
The pairing coherence length slowly decreases with increasing temperature, indicating a gradual quenching of pairing correlations, as is natural in finite systems.
%%%%%
In exotic nuclei close to drip lines, where the Fermi energy has positive values,
the correlation length has larger values and therefore the spatial correlation increases.

The quarteting coherence length describes correlations between proton and neutron pairs,
by overlapping their relative parts to the corresponding pp and nn components of 
the $\alpha$-particle wave function.
It has a similar behaviour, but with larger values on the nuclear surface. 
We evidenced the important role played by proton-neutron correlations by 
considering in addition the overlap with the pn component of the $\alpha$-particle
wave function. 
They change completely the behaviour of the quarteting
coherence length, namely the alpha correlation length has oscillating
values around the $\alpha$-particle geometrical radius. 
Its mean value $\approx$ 1.7 fm weakly depends on the nuclear mass.
The analysis of the two terms entering the definition of the coherence length
reveales the main difference between the pairing and quarteting cases.
It turns out that pairing correlations are larger inside nucleus,
while quarteting correlations are connected to the nuclear surface.

%\newpage
\acknowledgments

This work has been supported by the project PN-II-ID-PCE-2011-3-0092 
and NuPNET-SARFEN of the Romanian Ministry of Education and Research.

%\newpage

\end{document}